\documentstyle[12pt,psfig]{article}
\advance\textheight by \topskip
\newcommand{\ba}         {\begin{array}}
\newcommand{\bc}         {\begin{center}}
\newcommand{\be}         {\begin{equation}}
\newcommand{\bea}        {\begin{eqnarray}}
\newcommand{\bib}        {\bibitem}
\newcommand{\bmp}[2]     {\begin{minipage}[#1]{#2}}
\newcommand{\bra}[1]     {\left.\langle#1\right|}
\newcommand{\cf}         {{\it cf.~}}

\newcommand{\ea}         {\end{array}}
\newcommand{\ec}         {\end{center}}
\newcommand{\ee}         {\end{equation}}
\newcommand{\eea}        {\end{eqnarray}}
\newcommand{\eg}         {{\it e.\,g.~}}
\newcommand{\emp}        {\end{minipage}}
\newcommand{\eps}        {\varepsilon}
\newcommand{\expv}[1]    {\langle#1\rangle}
\newcommand{\gapp}       {\mathrel{\rlap {\raise.5ex\hbox{$>$}}
                          {\lower.5ex\hbox{$\sim$}}}}

\newcommand{\hvect}[1]   {\hat{\mbox{\boldmath$#1$}}}
\newcommand{\ie}         {{\it i.\,e.~}}
\newcommand{\ket}[1]     {\left|#1\rangle\right.}
\newcommand{\la}         {\langle}
\newcommand{\lapp}       {\mathrel{\rlap{\raise.5ex\hbox{$<$}}
                          {\lower.5ex\hbox{$\sim$}}}}

\newcommand{\ra}         {\rangle}
\newcommand{\rslash} [1] {\mbox{#1\hspace{-7pt}/}}

\newcommand{\sla}        {\kern -5.4pt /}
\newcommand{\sm}[1]      {\SS #1\TS}
\newcommand{\sprod}[2]   {\vect{#1}\!\cdot\!\vect{#2}}
\newcommand{\SS}         {\scriptstyle}
\newcommand{\SSS}        {\scriptscriptstyle}

\newcommand{\TS}         {\textstyle}

\newcommand{\vect}[1]    {\mbox{\boldmath$#1$}}

\newcommand{\cA}{{\cal A}}

\newcommand{\cI}{{\cal I}}

\newcommand{\cO}{{\cal O}}

\newcommand{\JP}     [1]{{\it J.\,Phys.\ }                     {\bf #1}}

\newcommand{\NC}     [1]{{\it Nuovo Cim.\ }                    {\bf #1}}
\newcommand{\NP}     [1]{{\it Nucl.\,Phys.\ }                  {\bf #1}}
\newcommand{\PL}     [1]{{\it Phys.\,Lett.\ }                  {\bf #1}}
\newcommand{\PR}     [1]{{\it Phys.\,Rep.\ }                   {\bf #1}}
\newcommand{\PRev}   [1]{{\it Phys.\,Rev.\ }                   {\bf #1}}

\newcommand{\ProcRS} [1]{{\it Proc.\,R.\,Soc.\ }               {\bf #1}}
\newcommand{\ProgPNP}[1]{{\it Prog.\,Part.\,and Nucl.\,Phys.\ }{\bf #1}}
\newcommand{\RMP}    [1]{{\it Rev.\,Mod.\,Phys.\ }             {\bf #1}}

\newcommand{\ZP}     [1]{{\it Z.\,Phys.\ }                     {\bf #1}}

\begin{document}
\begin{titlepage}
\begin{flushright}
FZR-55\\
nucl-th/9409011
\end{flushright}
\bc
{\Large\bf Pushing and Cranking Corrections\\[2mm]
to the Meson Fields\\[4mm]
of the Bosonized Nambu \& Jona-Lasinio Model}
\ec
\vskip\baselineskip \vskip\baselineskip
\bc
{\sc M.\,Schleif$^{a,b}$ and R.\,W\"unsch$^a$}

\bigskip \bigskip

$^a$Institut f\"ur Kern- und Hadronenphysik, Forschungszentrum
Rossendorf e.\,V.,\\
Postfach 51\,01\,19, D-01314 Dresden, Germany\\[2mm]
$^b$Institut f\"ur Theoretische Physik, Technische Universit\"at Dresden\\
Mommsenstr.\,13, D-01062 Dresden, Germany
\ec
\vskip\baselineskip
\vskip\baselineskip
\vskip\baselineskip

\begin{abstract}
We study the effect of center-of-mass motion and rotational corrections
on hedgehog meson fields in the bosonized two-flavor Nambu \& Jona-Lasinio
model.
To remove the spurious motion and to restore good spin and isospin we consider
a boosted and rotating soliton instead a static soliton at rest.
Modified meson fields are obtained by minimizing a corrected effective energy
functional.
The importance of the modification is estimated by evaluating expectation
values of several observables.

Stabile solitonic configurations are obtained for $M\gapp$ 300\,MeV,
while static solitons exists for $M\gapp$ 350\,MeV only.
Despite the considerable size of the energy corrections
(30-50\% of the soliton energy) the main features of the static
soliton are preserved.
Modified meson profiles are somewhat narrower than static ones and have a
different asymptotic behavior, which depends on the isospin quantum number.
The modifications increase with increasing constituent quark mass.
The valence-quark picture dominates up to very large constituent quark masses.

In the physically relevant mass region, the root-mean square radius of the
quark distribution is reduced by less than 10 percent.
The $\Delta$--nucleon mass-splitting is still weaker affected.
\end{abstract}
\end{titlepage}

\section{Introduction}

Chiral soliton models have proved to be a fruitful approach to the
description of nucleon structure. We consider a soliton which is defined by
the {\em effective action} of the bosonized
form \cite{Egu-76,Kle-76} of the Nambu \& Jona-Lasinio (NJL) model
\cite{Nam-61}.
In mean-field approximation, the mesonic field is treated on the classical
level (zero loop) and minimizes the Euklidean effective action.
Static classical meson fields are determined by an {\em effective energy}.
Quark fields are obtained by diagonalizing a hamiltonian with quark-meson
interaction.
Infinite sea-quark contributions have to be regularized. We apply Schwinger's
proper-time scheme \cite{Sch-51}.
Solitonic meson fields restricted to spherical hedgehog configurations and to
the chiral circle are uniquely determined by a {\em profile function}
$\Theta(r)$, which depends on the separation $r$ from the center of the
soliton.
Relating interaction strength, regularization parameter and current quark
mass to the experimental values of the weak pion-decay constant and the pion
mass, the constituent quark mass is the only free parameter in the model.
Solitonic hedgehog configurations have been obtained for constituent
quark masses $M\gapp$ 350\,MeV.
For recent reviews see refs.\,\cite{Vog-91,Kle-92,Hat-94,Mei-94}.

Even though the fundamental NJL lagrangian possesses translational symmetry,
the mean-field solution does not share this symmetry in general.  The soliton
state is fixed to a definite spatial point and is not an eigenstate of the
center-of-mass momentum operator $\vect{P}$.
So it happens that the expectation value  $\expv{\vect{P}}$
vanishes for symmetry reasons, while the energy of the center-of-mass motion
(CMM), which is proportional to $\expv{\vect{P}^2}$, has a finite value.
The soliton is affected by the spurious CMM, whose kinetic energy
is a part of the total soliton energy. This spurious motion smears
out the quark distribution and affects mean field, mass distribution and
other observables.
The restriction to hedgehog configurations creates another defect of
the soliton. Spin $\vect{J}$ and isospin $\vect{T}$ fail to be good quantum
numbers and the expectation values
$\expv{\vect{J}^2}$ and $\expv{\vect{T}^2}$ do not agree with
the experimental values of the nucleon or the $\Delta$ isobar.
As a result of the restriction to hedgehog configurations the soliton is
eigenstate of the grand spin $\vect{G}=\vect{J}+\vect{T}$ with eigenvalue
$G=0$, and one has
$\vect{J}=\vect{T}$ and $\expv{\vect{J}^2}=\expv{\vect{T}^2}$ automatically
\cite{Bir-86}.

To remove the spurious CMM exactly one would have to reduce
the number of degrees of freedom by introducing the
center-of-mass coordinate explicitly and describing the quarks by
intrinsic coordinates. Since the Dirac sea of quark states is involved
in the model such a procedure is not feasible. For similar reasons it is not
feasible to abandon the restriction to hedgehog configurations
and to care about good spin and isospin from the very beginning.
So one usually accepts the shortcomings of a symmetry-violating mean field
in hedgehog shape, calculates field configurations within the
restricted basis and tries to restore the correct values of
$\la\vect{P}\,^2\ra$, $\expv{\vect{J}^2}$ and $\expv{\vect{T}^2}$ afterwards.

Several approximations have been developed to remove CMM and correct
spin and isospin \cite{Bir-90}.
We apply the semiclassical {\em pushing} and {\em cranking} approaches
\cite{Rin-80}. Instead of a field configuration at rest we consider the
soliton in a frame boosted with the velocity $\vect{V}$ and rotating with
the angular frequency $\vect{\Omega}$ in isospace
\cite{Coh-86,Adk-83,Rei-89}.
Both quantities $\vect{V}$ and $\vect{\Omega}$ are Lagrange
multipliers and fixed such  that the correct values
$\expv{\vect{P}^2}=0$ and $\expv{\vect{T}^2}=T(T+1)$ are obtained.
The pushing approach is the translational analogue to the cranking which
itself is equivalent to a semiclassical approximation of the projection
method by Peierls and Yoccoz \cite{Rin-80,Pei-57,Gri-57,Bla-86}.

The energy of a boosted and rotating soliton differs from the energy of a
soliton at rest by the kinetic energy of the collective motions.
Both translational and rotational energy are described by inertial
parameters which depend on the field configuration.
Calculating the inertial parameters or other expectation values in
first-order perturbation approximation one uses meson fields which minimize
the static soliton energy.
The fields are assumed to be not affected by the collective motion
(rigid rotation).
In fact, the fields should be distorted by the relativistic boost and the
centrifugal forces. The neglect of this response is equivalent to the
{\em variation before projection} in the projection approach of symmetry
restoration.

Because of the considerable size of the energy corrections (see fig.\,1)
we improve this approach.
Instead of static meson and quark fields we use modified fields
defined by a corrected effective energy which includes the kinetic energy
of the collective motions.
This method is equivalent to the {\em variation after projection}.
Now the inertial parameters are allowed to respond
to the motion of the soliton (self-consistent pushing and cranking).
Since the rotational energy depends on the isospin quantum number $T$
one gets slightly different fields for nucleon and $\Delta$ isobar.

It is the aim of this paper to evaluate the modified meson fields and the
complementary quark fields.
In order to study size and importance of the field modifications we evaluate
expectation values of several observables. Possible modifications of the
observables themselves, as a consequence of the collective motion, are not
considered.

In sect.\,2, we define static soliton configurations and modify them by CMM
and rotational corrections.
The numerical procedure used for the calculation of the modified fields is
outlined and tested in sect.\,3. Modified fields and expectation values
are calculated in sect.\,4. Their deviations from the unmodified quantities
are interpreted on the basis of quasi-classical arguments.
Conclusions are drawn in sect.\,5.

\section{Static, pushed and cranked solitons}

We start from the static soliton obtained in the bosonized version of the
Nambu \& Jona-Lasinio model for temperature $T=0$ and finite chemical
potential $\mu$ for quarks. Let us restrict the model to the two light
quarks with a common current mass
$m$ and a chirally invariant combination of a scalar--isoscalar and a
pseudoscalar--isovector quark-quark interaction.
Introducing classical meson fields as the expectation values
$\sigma\sm{(}x\sm{)}\sim\la\bar{q}\sm{(}x\sm{)}\,q\sm{(}x\sm{)}\ra\;
\mbox{and}\;
\hat{\vect{\pi}}\sm{(}x\sm{)}\sim\la\bar{q}\sm{(}x\sm{)}\,
i\gamma_5\hat{\vect{\tau}}\,q\sm{(}x\sm{)}\ra$
of bilinear combinations of quark operators $q\sm{(}x\sm{)}$
with the vector $\hat{\vect{\tau}}$ of Pauli matrices
the system can be described by an effective Euklidean action
$\cA_{eff}\left[\sigma,\hat{\vect{\pi}}\right]$,
which is a functional of the mesonic fields.
Restricting the mesonic fields to static hedgehog configurations and to the
chiral circle they are
uniquely described by a single profile function $\Theta\sm{(}r\sm{)}$,
where $r$ is the distance from the center of the spherically symmetric
fields.
In the case of static meson fields the effective action is proportional to
the Euklidean time interval $\int\!d\tau$, and an effective energy
\be\label{Eeff}
E[\Theta]\sm{(}\mu\sm{)}\,\int\!d\tau\,=\,
\cA_{eff}\left[\Theta\right]\sm{(}\mu\sm{)}
\ee
can be defined.
The profile function $\Theta\sm{(}r\sm{)}$ minimizes the effective energy
(\ref{Eeff})
\be\label{Emin}
E\sm{(}\mu\sm{)}\,=\,
\min\limits_{\Theta\SSS(\SS r\SSS)}E[\Theta]\sm{(}\mu\sm{)}
\ee
and fulfills the equation of motion
\be\label{EOM}
\frac{\delta E\left[\Theta\right]\sm{(}\mu\sm{)}}{\delta\Theta\sm{(}r\sm{)}}
\,=\,0
\ee
for a given chemical potential $\mu$.
The effective energy (\ref{Eeff}) consists of a term $E^q$, which results from
the Fermion determinant, and of a purely mesonic contribution $E^m$
\be\label{Eeffqm}
E\sm{(}\mu\sm{)}\,=\,
E^q\sm{(}\mu\sm{)}+E^m.
\ee
Within the restrictions we have imposed on the mesonic fields
the latter reduces to
\be\label{Ebr}
E^m\,=\,E^{br}\,=\,
m\,f_\pi\frac{\lambda^2}{g}\int\!d^3\!\vect{r}\,
\left[1-\cos{\Theta\sm{(}r\sm{)}}\right],
\ee
which stems from the mass term in the original NJL Lagrangian and
breaks chiral symmetry explicitly.
The quark contribution to the effective energy is given by
\be\label{Fermidet}
E^q\sm{(}\mu\sm{)}\int\!d\tau\,=\,
-Sp\,Log\left(\rslash{D}_E-\mu\beta\right)
\ee
with the Euklidean Dirac operator
\be\label{DirEuk}
\rslash{D}_E\,=\,\beta\left(\frac{\partial}{\partial\tau}+h(\Theta)\right)
\ee
and the quark hamiltonian
\be\label{QHam}
h(\Theta)\,=\, \sprod{\alpha}{p}+
gf_\pi\,\beta\left(\cos{\Theta}+i\gamma_5\sin{\Theta}\,\hvect{\tau}\!\cdot\!
\hvect{r}\right).
\ee
The chemical potential $\mu$  was introduced in the fermion determinant
(\ref{Fermidet})
as a Lagrange multiplier in order to adjust the baryon number $B$.
The symbol $Sp$ includes functional trace $\left(\int\!d^4x_E\right)$ over
the Euklidean space-time $x_E\!=\!(\tau,\vect{r})$ and matrix trace over
Dirac $\left(tr_\gamma\right)$, isospin $\left(tr_\tau\right)$ and color
indices. The latter results in a factor $N_c\!=\!3$ due to
color symmetry
\be\label{SpDef}
Sp\,\cO\,\equiv\,
N_c\,tr_\gamma\,tr_\tau\int\!d^4x_E\bra{x_E}\cO\ket{x_E}.
\ee
The parameters $m$ and $f_\pi$ denote current quark mass and weak pion-decay
constant, respectively, and $\vect{\alpha}$, $\beta$ are Dirac matrices.
The vacuum state is defined by $\Theta\sm{(}r\sm{)}\!\equiv\!0$ and marked
by the upper index $V$. The vacuum hamiltonian
$h^V\!\equiv\!h(\Theta\!\equiv\!0)$ describes quarks
with a constituent mass $M\!=\!gf_\pi$.

In the following we consider $B\!=\!1$ configurations and assume the
chemical potential to be fixed such that
\be\label{Bnumber}
B\sm{(}\mu\sm{)}\,=\,1
\ee
is fulfilled. Expectation values $\expv{\cO}$ as well as the energies
(\ref{Eeffqm}, \ref{Ebr}, \ref{Fermidet}) are defined
in accordance with the conditions (\ref{EOM}) and (\ref{Bnumber})
\be\label{expv}
\expv{\cO}\,\equiv\,\expv{\cO}[\Theta]\sm{(}\mu\sm{)}
\bigg|_{\frac{\delta E[\Theta]}{\delta\Theta\SSS(\SS r\SSS)}\SS=0,\,
B\SSS(\SS\mu\SSS)\SS=1}.
\ee

The quark energy (\ref{Fermidet}) can be split into a {\em valence
contribution} $E_{val}$ and a term $E_{sea}$ which is called the
{\em Dirac-sea contribution}
\be\label{Eeffq}
E^q\,=\,E_{val}+E_{sea}.
\ee
We should mention that the notation Dirac-sea contribution is conventional
but does not agree with the Dirac sea as the continuum of states with
negative energy. Here, all the levels, both continuous and discrete,
contribute to the Dirac sea.
Sometimes \cite{Wak-91} it is denotes as {\em vacuum polarization}
but this may be misleading too.

The sea contribution is UV divergent and is regularized within
Schwinger's proper-time scheme \cite{Sch-51} introducing an regulator
function $R_E(\eps,\Lambda)$ with an additional cut-off parameter $\Lambda$
\cite{Rei-88,Mei-89}.
Valence and regularized sea energy can be expressed by the eigenvalues
$\eps_\alpha$ of the hamiltonian (\ref{QHam})
\be\label{Eval}
E_{val}\,=\,N_c\!\sum_{0\le\eps_\alpha\le\mu}\!\eps_\alpha,
\ee
and
\be\label{Eseareg}
E_{sea}^{Reg}\,\equiv\,E^{Reg}(\mu\!=\!0)-E^{V,Reg}(\mu\!=\!0)\,=\,
-\frac{N_c}{2}\sum\limits_\alpha\left[
R_E(\eps_\alpha,\Lambda)\left|\eps_\alpha\right|-
R_E(\eps_\alpha^V,\Lambda)\left|\eps_\alpha^V\right|\,\right],
\ee
where we have subtracted the vacuum energy $E^V$ with the eigenvalues
$\eps^V_\alpha$ of $h^V$ from the sea contribution. A similar representation
can be found for other observables \cite{Mei-94}.
The parameters $\Lambda$, $g$, $\lambda$ and $m$ are fixed
in the mesonic as well as the vacuum sector leaving the constituent quark
mass $M$
as the only parameter for the baryonic sector \cite{Mei-94,Mei-91a}.

Generally a profile function $\Theta\sm{(}r\sm{)}$, which minimizes the
effective energy (\ref{Eeffqm}), is numerically determined.
In foregoing papers \cite{Rei-88,Wue-94} we calculated
$\Theta\sm{(}r\sm{)}$ iteratively using the equation of motion (\ref{EOM})
and diagonalizing the quark hamiltonian (\ref{QHam}) in
the discrete basis introduced in ref.\,\cite{Kah-84}. We found solitonic,
\ie spatially restricted
configuration for $M\gapp$ 350\,MeV and calculated their parameters
\cite{Wue-94,Goe-91a}.

To remove the spurious translational and rotational degrees of freedom
approximately and to equip the soliton with correct spin and isospin we
apply the semiclassical pushing and cranking approaches \cite{Rin-80}.
Instead of a static soliton at rest we consider a
soliton pushed with the velocity $\vect{V}$ and cranked with the angular
velocity $\vect{\Omega}$ in isospace.
Expanding the corresponding effective action up to second order in
$\vect{V}$ and
$\vect{\Omega}$ one gets a corrected effective energy \cite{Pob-92}
\be\label{ECORR}
E^T_{CORR}\,=\,E-E_{CMM}+E^T_{ROT},
\ee
where $E$ is the static effective action defined in eq.\,(\ref{Eeff}).
If one fixes the parameter $\vect{V}^2$ such that the boosted soliton has
$\expv{\vect{P}^2}=0$ one gets
\be\label{ECMM}
E_{CMM}\,=\,
\frac{\expv{\vect{P}^2}}{2E},
\ee
where the inertial parameter $E$ is identical \cite{Pob-92} with the
effective energy (\ref{Eeff}). The energy correction (\ref{ECMM})
describes the spurious translational
energy contained in the soliton energy (\ref{Eeff}). It appears from the
translational-symmetry violating mean field and has to be subtracted.

The rotational correction $E^T_{ROT}$ is determined by quantizing the
rotational degree of freedom semiclassically according to
\be\label{rotquant}
\expv{\vect{T}^2}\,=\,\vect{\Omega}^2\,\cI^2\qquad\Longrightarrow\qquad
|\vect{\Omega}|\,=\,\frac{\sqrt{T(T+1)}}\cI,
\ee
and one gets \cite{Pob-92}
\be
\label{EROT}
E_{ROT}^T\,=\,\frac{T(T+1)}{2\cI}-\frac{9/4}{2\cI}\,=\,
         \left\{ \begin{array}{r@{\quad for \quad}c}
         - \frac{3}{4\cI} & \mbox{nucleons} \\[2mm]
           \frac{3}{4\cI} & \Delta \mbox{ isobars}
         \end{array} \right.\:.
\ee
The first term describes the energy of a rotor in isospace with isospin
$T$ and moment of inertia $\cI$.
The second term corresponds to the band-head energy \cite{Rin-80} and
accounts for the finite
value of $\expv{\vect{T}^2}$ in the static hedgehog configuration.
It results from the valence quarks only and describes the spurious
rotational energy of the quarks interacting with a meson field of
hedgehog shape, which  violates isospin symmetry.
Both kinds of spurious energies, translational
and rotational, are independent of the number $N_c$ of colors, while the
energy of the rotor is proportional to $N_c^{-1}$.

The inertial parameters $E$ and $\cI$ in eqs.\,(\ref{ECMM}) and (\ref{EROT})
depend functionally on the meson profile $\Theta\sm{(}r\sm{)}$.
They can be expressed by the eigenvalues of the quark hamiltonian (\ref{QHam})
and by the matrix elements of the isospin operator $\hat{\vect{\tau}}$
with the eigenfunctions of the quark hamiltonian. The
moment of inertia $\cI$ is given by a regulated {\em Inglis formula}
\cite{Rei-89}. In
\begin{figure}[h]

\vspace*{-2.8cm}

\begin{minipage}{8cm}
\hspace*{-2mm}
\unitlength1cm
\centerline{
\psfig{file=fig1.ps,height=12cm,width=14cm,angle=90}}
\end{minipage}
\hfill
\begin{minipage}{6.cm}
{\em Figure 1}:
Spurious center-of-mass energy $E_{CMM}$ (\ref{ECMM}) and absolute value
$|E_{ROT}|$ of the rotational energy (\ref{EROT}) calculated for
uncorrected meson profiles $\Theta\sm{(}r\sm{)}$, in dependence on the
constituent quark mass $M$.
\end{minipage}
\end{figure}

\vspace*{-1cm}\noindent
fig.\,1, we display the energy corrections calculated for static fields,
\ie for mesonic fields which minimize the effective energy (\ref{Eeff})
and for the corresponding quark fields. The spurious part of the rotational
energy is given by $\frac{3}{2}|E_{ROT}|$.

Within the static approximation, center-of-mass energy (\ref{ECMM}) amounts
to 15--30 percent in the physically relevant region of small constituent
quark masses (350\,MeV $\le M\le$ 500\,MeV)
and reaches 50 percent of the total soliton energy of roughly 1240\,MeV for
$M$=1000\,MeV.
The rotational corrections (\ref{EROT}) cancel out partly. On the whole they
amount to roughly half of the CMM correction, for small constituent quark
masses.

The increase of the spurious energies with increasing mass parameter $M$
can be understood as follows.
The main contribution to the center-of-mass energy and to the moment of
inertia is recruited from the valence quarks, which are confined within a
small volume.
Increasing the constituent quark mass this volume shrinks and the confined
quarks get a larger kinetic energy due to Heisenberg's principle.
Increasing kinetic energy involves increasing center-of-mass energy,
which is a fraction of it. On the other hand, a more compact mass
distribution has a
smaller moment of inertia, which is in the denominator of the rotational
energy (\ref{EROT}).
For constituent masses $M\gapp$ 750\,MeV, the attraction between the quarks
is so strong
that the valence level dives into the negative-energy region.
Now there is obviously no more valence contribution to any expectation value.
The former valence level, however, continues to be confined
and to give the dominating contribution, now as a member of the Dirac sea.

The energy corrections in fig.\,1 are in fair agreement with a calculation
in the Gell-Mann--Levi
\cite{Gel-60} model using the Peierls-Yoccoz projection \cite{Neu-92}.
Comparing both results one has to take into account that, in the bosonized
NJL model,
the mass $m_\sigma$ of the $\sigma$ meson is related to the constituent
quark mass $M$
via $m_\sigma^2=4M^2+m_\pi^2$, while it is a free parameter in the
Gell-Mann--Levi model.
The spurious part of the energy corrections can be calculated also as
quantum corrections to the
mean-field approximation \cite{Wei-94}. The evaluated corrections
to the soliton energy due to rotational zero modes are in good agreement
with our spurious rotational energies, while the translational quantum
fluctuations yield only half of our CMM corrections.
The difference can be explained by the truncation of the meson modes
in ref.\,\cite{Wei-94}.

The considerable size of the corrections led us to go beyond the first-order
consideration and to determine CMM and rotational corrections
self-consistently.
For that aim we determine modified profile functions
$\Theta^T_{mod}\sm{(}r\sm{)}$
which minimize the corrected effective energy (\ref{ECORR}) instead of the
effective energy $E$ (\ref{Eeff})
\be\label{ECORRmin}
E_{CORR,mod}^T\,\equiv\,E^T_{CORR}[\Theta^T_{mod}]\,=\,
\min\limits_{\Theta\SSS(\SS r\SSS)}E^T_{CORR}[\Theta].
\ee
The profiles $\Theta\sm{(}r\sm{)}$ and $\Theta^T_{mod}\sm{(}r\sm{)}$ deviate
 from each other, since the correction terms (\ref{ECMM}) and (\ref{EROT})
depend functionally on the meson profile $\Theta\sm{(}r\sm{)}$.
Modified expectation values are defined by
\be\label{expvmod}
\expv{\cO}_{mod}\,\equiv\,\expv{\cO}[\Theta]\sm{(}\mu\sm{)}
\bigg|_{\Theta=\Theta^T_{CORR},\;B\SSS(\SS\mu\SSS)\SS=1}.
\ee
Modified profiles depend on isospin, since the rotational correction
(\ref{EROT}) depends on the isospin quantum number $T$. Meson profiles for
nucleons are different from profiles for $\Delta$ isobars.
This may be the reason for different expectation values
(\eg the baryon root-mean square radius), which are identical otherwise.

Modified profile functions fulfill a modified equation of motion
\be\label{EOMmod}
\left.\frac{\delta E_{CORR}^T[\Theta]}
{\delta\Theta(r)}\right|_{\Theta=\Theta^T_{mod}}\,=\hspace*{105mm}
\ee
\[\hspace*{35mm}
\left.\frac{\delta E[\Theta]}
{\delta\Theta(r)}\right|_{\Theta=\Theta^T_{mod}}\!\!-
\left.\frac{\delta E_{CMM}[\Theta]}
{\delta\Theta(r)}\right|_{\Theta=\Theta^T_{mod}}\!\!+
\left.\frac{\delta E_{ROT}^T[\Theta]}
{\delta\Theta(r)}\right|_{\Theta=\Theta^T_{mod}}=\;0.
\]
This equation is much more complicated than the
unmodified equation (\ref{EOM}) since the correction terms $E_{CMM}$ and
$E_{ROT}^T$ depend on the profile function via the inertial parameter $\cI$
and the expectation value $\expv{\vect{P}^2}$.
It is not feasible to determine $\Theta^T_{mod}\sm{(}r\sm{)}$ iteratively.
We use a N-parameter representation of the profile function
and minimize the corrected energy functional (\ref{ECORR}) numerically.
The procedure will be explained in the next section.

\section{Spline representation of the meson profile with predetermined
asymptotic behavior}

With the aim to minimize the corrected soliton energy (\ref{ECORR})
numerically we parametrize the profile function.
The asymptotic behavior of the uncorrected profile at small and large
separations $r$ can be determined analytically \cite{Bir-90,Mei-91a}
\be\label{Theasy}
\Theta\sm{(}r\sm{)}\,\longrightarrow\,\left\{\begin{array}{lll}
-\pi(1-ar) & \mbox{for} & r\to0\\[2mm]
-b\frac{1+m_\pi r}{r^2}e^{-m_\pi r} & \mbox{for} & r\to\infty
\end{array}\right.,
\ee
where $m_\pi$ is the pion rest mass and $a$, $b$ are parameters.
In ref.\,\cite{Wue-94} we introduced a reference profile which interpolates
smoothly between the two asymptotic pieces.
Now we parametrize the profile by $N$ values
$\Theta_i\equiv\Theta\sm{(}r_i\sm{)}\;(i=1,\ldots,N)$ at the points
$r_1,\ldots,r_N$ (knots) in the intermediate region.
The values $\Theta_1$ and $\Theta_N$ determine the parameters $a$ and $b$ in
eq.\,(\ref{Theasy}).
A spline interpolation determines the profile between the knots and
eq.\,(\ref{Theasy}) is used for $r\le r_1$ and $r\ge r_N$.
Within this representation, the effective energy is an ordinary functions
of the parameters
$\Theta_i\;(i=1,\ldots,N)$ and can be minimized by standard methods.
This procedure can be used for the static as well as the corrected energy
functional.
The method was tested for the static energy (\ref{Eeffqm}) first.
Here we can compare the result of the minimization (\ref{Emin}) with the
iterative solution of the equation of motion (\ref{EOM}).
We found that $N$=7 knots distributed around the average radius $R$ defined by
\be\label{Rave}
\Theta(R)\,=\,
\frac{1}{2}\left[\Theta(0)-\Theta(\infty)\right]\,=\,-\frac{\pi}{2}
\ee
are sufficient for an accurate reproduction of the meson profile.
The location of the knots is illustrated in the upper part of fig.\,2.
The first knots $r_1$ and $r_2$ are situated at $R/2$ and $R$, respectively.
The last knot $r_N$ is located at a radius $R^{asy}$. It marks the point
where the meson profile is sufficiently well described by the asymptotic
formula (\ref{Theasy}).
Empirically we found $R^{asy}\approx5R$.
The other knots lie between $R$ and $R^{asy}$.

The accuracy of such a spline interpolation with predetermined asymptotic
behavior was numerically checked. First we reproduced self-consistently
determined profiles by means of their values at the 7 selected knots.
The central part of fig.\,2 shows the deviation $\delta\Theta$ of the
interpolation from the original for $M$=500\,MeV.
The same test was performed for the profiles in the whole mass region
under investigation. The deviations did not exceeded $0.25^o$.
After testing the 7-knots spline reproductions of a given
profile functions we compared the iteratively determined profile function
with the result of the numerical minimization of the energy functional
(\ref{Emin}) using the 7-knots spline representation.
For $M$=500\,MeV, the difference between the two results is shown in the
lower part of fig.\,2.
The deviations are less than $0.35^o$ in the whole mass region.

Finally we tested the sensitivity of expectation values with respect to our
approximations. We compared expectation values calculated for iteratively
determined profiles with their values for spline-reproduced profiles and for
profiles obtained by means of the minimization procedure. Using the 7-knots
spline valence- and sea-quark energies vary by not more than 1\,MeV.
The mesonic energy (\ref{Ebr}) is more sensitive to details of the meson
profile and varies up to 2\,MeV. The variation of the root-mean square
radius is limited to 0.02\,fm.
Apart from constituent quark masses around 350\,MeV, where the modifications
in the meson profile are very small, the effect of CMM and rotational
corrections is remarkably larger
than the uncertainties in the numerical procedure. So we conclude that the
accuracy of the spline representation of the profile function with
predetermined
asymptotic behavior and of the numerical minimization procedure is adequate
to the pretensions of the whole model.
A larger number of knots increases the calculation time noticeably without
improving the accuracy considerably.

Before applying the method to the corrected energy functional (\ref{ECORR})
we have to consider the asymptotic behavior for $r\to\infty$ for a rotating
soliton, which deviates from eq.\,(\ref{Theasy}) due to the action of the
centrifugal forces.
In the asymptotic region, the isovector $\hat{\vect{\pi}}$ field of the
bosonized NJL model has to fulfill the same differential equation as the
corresponding field
in the Skyrme model \cite{Sky-61} and in the chiral sigma model of Gell-Mann
and Levi \cite{Gel-60} as well.
So we can exploit the insights obtained within these models.
As shown \eg in refs.\cite{Bla-88,Pos-89,Dor-94} the $\hat{\vect{\pi}}$ field
of a rotating soliton has the same asymptotic behavior (\ref{Theasy}) as the
static pro-

\vspace*{-17mm}

\hspace*{-37mm}
\begin{minipage}[t]{10cm}
\mbox{
\psfig{file=fig2a.ps,height=10cm,width=13cm,angle=90}}

\vspace*{-4.6cm}

\mbox{
\psfig{file=fig2b.ps,height=7cm,width=13cm,angle=90}}

\vspace*{-38.0mm}

\mbox{
\psfig{file=fig2c.ps,height=7cm,width=13cm,angle=90}}
\end{minipage}
\hfill
\begin{minipage}[t]{6.cm}

\vspace*{-68mm}

{\em Figure} 2: Spline representation with predetermined asymptotic behavior
and 7 knots.\\
{\em Upper part}: Shape of the profile function $\Theta\sm{(}r\sm{)}$ for
$M$=500\,MeV
and position of the knots indicated by arrows.\\
{\em Central part}:
Deviation of the iteratively calculated profile from its
7-knot spline representation with predetermined asymptotic behavior.\\
{\em Lower Part}:
Differences between profiles determined iteratively by means
of the equation of motion and profiles obtained by minimizing the energy
directly using the spline interpolation.
\end{minipage}

\vspace{-2mm}

\noindent
file function, however, in the components perpendicular to the
rotational axis, the pion rest mass $m_\pi$ has to be replaced by the
modified pion mass
\be\label{mpitilde}
\tilde{m}_\pi\,=\,\sqrt{m_\pi^2-\vect{\Omega}^2}.
\ee
For rotational frequencies $|\vect{\Omega}|$ which are comparable with the
pion rest mass,
the rotationally improved soliton is much more diffused than the static one
($\tilde{m}_\pi<m_\pi$).
If the rotational frequency, which is necessary to produce the correct
expectation values of spin and isospin, is larger than the pion rest mass
the situation
changes dramatically. Instead of vanishing exponentially the components of
$\hat{\vect{\pi}}$ perpendicular to $\vect{\Omega}$ start to oscillate.
This corresponds to the emission of pions and may be used for the description
of the decay of $\Delta$ isobars.
Already for frequencies $|\vect{\Omega}|<m_\pi$ the shape of the rotationally
improved soliton deviates from the hedgehog. It is neither spherically
symmetric nor
is the direction of the isovector $\hat{\vect{\pi}}$ field aligned
with the direction of vector $\vect{r}$.
Such fields can not be characterized by a single profile function
$\Theta\sm{(}r\sm{)}$
and must be treated on a three-dimensional grid \cite{Pos-89}.
Numerical tests have shown that quark observables are not very sensitive to
the asymptotic
behavior of the meson fields, in particular, if they are dominated
by the valence contribution. So we retain the hedgehog structure with the
asymptotics defined in eq.\,(\ref{Theasy}).
In this way the effect of the centrifugal force is spherically symmetric
spread over all directions. At large separations $r\le r_N$, the rotational
correction may change the parameter $b$ in the asymptotics only.
The $\Delta$ isobar we are considering is an artificially stabilized particle.

The manipulation of the asymptotic behavior is a well known procedure in
nuclear physics.
Resonance states, which emit particles and should be described by oscillating
wave functions can successfully be modeled by harmonic oscillator wave
functions which decrease exponentially. These wave functions reproduce most
of the properties
of the resonance state, in particular such properties which are determined by
the interior of the wave function. Of course, decay properties, which depend
essentially on the asymptotic behavior, cannot be described within this
approach.

\section{Profile functions and expectation values
modified by pushing and cranking corrections}

Fig.\,3 illustrates the general features of the modification in the profile
function
caused by pushing and cranking corrections in the effective energy functional.
To get a clear effect we chose the relatively large constituent quark mass
of 600\,MeV.
For smaller masses the effect is analogous, only smaller in size.
The asymptotic region starts outside the figure at $r=R^{asy}\approx6M^{-1}$.
The inner linear part is affected by the center-of-mass motion. The boost
applied to the soliton is orientated such that a part
of the kinetic energy of the quarks, mainly of the valence quarks, is
removed.
The size of the soliton is balanced by the attraction between the quarks
mediated by
the meson fields, and by the motion of quarks inflating the soliton.
Reducing the latter the soliton shrinks. For masses $M\lapp$ 350\,MeV the
attractive forces are not strong enough to produce a stabile soliton.
The reduction of the internal kinetic energy stabilizes the soliton.
CMM corrected solitons are already stable for $M\gapp$ 300\,MeV.

\vspace*{-25mm}

\hspace*{-40mm}
\begin{minipage}[b]{7.5cm}
\mbox{
\psfig{file=fig3.ps,height=12cm,width=14cm,angle=90}}
\end{minipage}
\hfill
\begin{minipage}[b]{5cm}
{\em Figure} 3:
Profile function $\Theta\sm{(}r\sm{)}$ of the static soliton
({\em broken line}) in comparison with the modified profiles of nucleon and
$\Delta$ isobar ({\em full lines}) for the constituent quark mass
$M$=600\,MeV.

\vspace*{40mm}
\end{minipage}

\vspace*{-10mm}

While the CMM correction is the same for nucleons and $\Delta$ isobars
the rotational correction has different sign. To get a $\Delta$ isobar one
has to add rotational energy to the soliton and the centrifugal force
presses the quarks
outwards. The meson field, which is produced by the quarks, follows this trend
($b_\Delta>b_{static}$). The opposite is true for the nucleon.
To get a particle with (iso)spin $\frac{1}{2}$ one has to take out a certain
amount of rotational energy and the resulting profile function tends to zero
more rapidly ($b_N<b_{static}$). Of course, these are classical
considerations, but they are adequate to the semiclassical way of
calculating the energy corrections.

Changing the constituent quark mass $M$ we can study the dependence of the
modifica-
\begin{minipage}[t]{6.5cm}
tions on the size of the correction terms (\cf fig.\,1).
First let us consider the CMM correction separately.
Though the absolute size of the correction is not the crucial quantity the
modifications grow with increasing correction term, \ie with increasing
constituent quark mass.
The effect is rather controlled by the variation of the correction with the
shape of the profile in comparison with the variation of static soliton energy
(see eq.\,(\ref{EOMmod})). A larger correction term gives the variation a
larger weight. A correction which is entirely independent of the meson
profile does not at all modify the profile. For the static profile the
static energy $E$ is minimal.
Starting from this profile $\Theta\sm{(}r\sm{)}$ is varied until the
increase of static energy exceeds the loss of CMM and rotational energy.

\hspace{2ex} In the following we compare expectation values calculated for
3 kinds of particles: the static soliton, the nucleon and the $\Delta$
isobar. Each of them have been obtained with the corresponding meson
profile and the resulting quark field.

\hspace{2ex} First let us consider the corrected soliton energy (\ref{ECORR})
displayed in fig.\,5.
The difference between bro-
\end{minipage}
\hfill
\begin{minipage}[t]{8cm}
\vspace*{-20mm}
\hspace*{-35mm}
\begin{minipage}[t]{10cm}
\mbox{
\psfig{file=fig4a.ps,height=8cm,width=13cm,angle=90}}

\vspace*{-4.57cm}

\mbox{
\psfig{file=fig4b.ps,height=8cm,width=13cm,angle=90}}

\vspace*{-4.78cm}

\mbox{
\psfig{file=fig4c.ps,height=87.56mm,width=13cm,angle=90}}
\end{minipage}

\vspace*{-5mm}

\hspace*{5mm}\noindent
\begin{minipage}{7.5cm}
{\em Figure} 4:
Profile functions $\Theta\sm{(}r\sm{)}$ of the static soliton
({\em broken lines})
in comparison with the corresponding CMM corrected profiles ({\em full lines})
for 3 values of the constituent quark mass $M$.
\end{minipage}
\end{minipage}

\noindent
ken and full lines represents the gain of energy due to the change
 from $\Theta\sm{(}r\sm{)}$ to $\Theta^T_{mod}\sm{(}r\sm{)}$.
Agreement with the experimental nucleon mass is obtained for small
constituent quark masses.
The mass of the $\Delta$ isobar is underestimated in the whole region of
constituent quark masses.

Now let us study the various contributions to the total soliton energy which
turn out to be more sensitive to changes in the meson profile.
Fig.\,6 shows valence- and sea-quark contributions, and the small mesonic
contribution as well. For the modified profile, the dependence of both
valence and sea energy on the constituent quark mass is clearly weaker
than for the static soliton.
The energy of the valence quarks is mainly determined by the size $R$ of the
meson profile and by the meson-quark coupling-constant $g$,
which is proportional to the constituent quark mass $M$.
Increasing the constituent quark mass the meson potential gets deeper and the
energy of the valence quark decreases.
As shown in ref.\,\cite{Wue-94} the self-consistent size of the unmodified
meson profile is nearly independent of the

\vspace*{-25mm}

\noindent
\begin{minipage}[b]{5cm}
{\em Figure} 5:
CMM and rotationally corrected nucleon and $\Delta$ energies (\ref{ECORR})
calculated for the static soliton ({\em broken lines}) in comparison with
the same energies for modified profile functions ({\em full lines}).\\[35mm]
\end{minipage}
\hfill
\hspace*{-20mm}
\begin{minipage}[b]{12.cm}
\mbox{
\psfig{file=fig5.ps,height=12cm,width=14cm,angle=90}}
\end{minipage}

\vspace*{-23mm}

\hspace*{-40mm}
\begin{minipage}[b]{12.cm}
\vspace*{-25mm}
\mbox{
\psfig{file=fig6a.ps,height=102.4mm,width=14cm,angle=90}}

\vspace*{-61.7mm}

\mbox{
\psfig{file=fig6b.ps,height=11cm,width=14cm,angle=90}}
\end{minipage}
\hfill
\begin{minipage}[b]{5cm}
{\em Figure} 6: Valence (\ref{Eval}), sea-quark (\ref{Eseareg}) and mesonic
contributions (\ref{Ebr}) to the total soliton energy for the static soliton
configuration ({\em broken lines}) in comparison with the
corresponding energies for the modified meson profiles ({\em full lines}).\\
In the {\em upper part}, the profile $\Theta_{mod}^{T=1/2}$ modified for
nucleons was used. The lower part shows the same energies for meson profiles
$\Theta_{mod}^{T=3/2}$ modified for the $\Delta$ isobar.

\vspace*{30mm}

\end{minipage}

\vspace*{-13mm}
\noindent
constituent quark mass.
The shrinking of the modified profile counteracts the decrease of the
valence level and lowers the slope of $E_{val}$ in dependence on $M$.
It even prevents the valence level from leaving the valence-energy region
$[0,\mu\,]$, what happens at $M\approx$ 750\,MeV for the static soliton.
The sea energy is affected by the regularization procedure.
When the profile shrinks some of the sea levels leave the energy region
taken into account for a fixed regularization parameter $\Lambda$.
This reduces the increase of sea energy in comparison with the unmodified
case.

\vspace*{-31mm}
\noindent
\begin{minipage}[b]{5.2cm}
\vspace*{15mm}
{\em Figure} 7:
Mesonic energy (\ref{Ebr}) calculated for uncorrected profiles
$\Theta\sm{(}r\sm{)}$ ({\em broken line}) in comparison with modified
profiles ({\em full lines}) for the nucleon and the $\Delta$ isobar,
in dependence on the constituent quark mass $M$.
\\[30mm]
\end{minipage}
\hfill
\hspace*{-20mm}
\begin{minipage}[b]{12.cm}
\mbox{
\psfig{file=fig7.ps,height=12cm,width=14cm,angle=90}}
\end{minipage}

\vspace{-13mm}

Restricted to the chiral circle the energy $E^m$ of the meson field is
reduced to the term $E^{br}$ (\ref{Ebr}). This is a rather small quantity
and considered in fig.\,7 separately.
As explained in ref.\,\cite{Mei-91a} it is related to the nuclear sigma
commutator.
Apart from very small constituent quark masses the modifications in the meson
profile decrease the meson energy noticeably. Since it has only a small share
in the total soliton energy it is not relevant for the stability of the
soliton.

\vspace*{-26mm}
\hspace*{-40mm}
\begin{minipage}[b]{11cm}
\mbox{
\psfig{file=fig8.ps,height=12cm,width=14cm,angle=90}}
\end{minipage}
\hfill
\begin{minipage}[b]{5.5cm}
{\em Figure} 8:
Root-mean square radius (\ref{Rms}) of the isoscalar density distribution
calculated for unmodified ({\em broken line}) and modified profiles
({\em full lines}) in dependence on the constituent quark mass $M$\\[35mm]
\end{minipage}

\vspace*{-12mm}
Another quantity characterizing a soliton is the root-mean square
(r.\,m.\,s.\,) radius of the isoscalar mass distribution. We consider
\be\label{Rms}
\bar{R}\,\equiv\,\left\langle R^2\right\rangle^\frac{1}{2}\,=\,
\left[ \int\!d^3\vect{r}\,r^2\varrho(\vect{r}) \right]^{\frac{1}{2}},
\ee
where $\rho\sm{(}\vect{r}\sm{)}$ is the isoscalar baryon density. It is
displayed in fig.\,8 for unmodified and modified meson profiles.
The r.\,m.\,s.~radius is dominated by
the contribution of the valence quark (see \eg ref.\cite{Wue-94}).
The valence quarks are localized in the neighborhood of the center of the
soliton and hence hardly affected by rotational corrections.
The main correction stems from the center-of-mass motion
and decreases the radius slightly in accordance with the smaller size of the
modified meson profile. This effect is independent of the isospin quantum
number and results in identical corrections for nucleons and $\Delta$
isobars.
Rotational corrections affect only the loosely bound valence quarks for
$M\lapp$ 400\,MeV, which reach to larger separations from the center.
Here the faster rotating $\Delta$ isobar has a remarkably larger
r.\,m.\,s.~radius.
The experimental nucleon radius is reached for $M\approx$ 350\,MeV, where
the difference between
nucleon and $\Delta$ isobar is not very pronounced.

\hspace*{-40mm}
\begin{minipage}[b]{12.cm}
\vspace*{-26mm}
\mbox{
\psfig{file=fig9a.ps,height=11cm,width=14cm,angle=90}}

\vspace*{-63.9mm}

\mbox{
\psfig{file=fig9b.ps,height=11cm,width=14cm,angle=90}}
\end{minipage}
\hfill
\begin{minipage}[b]{5.5cm}
{\em Figure} 9:
Moment of inertia $\cI$ ({\em upper part}) and resulting
$\Delta$--nucleon mass-splitting $\Delta M_{\Delta N}$ ({\em lower part})
calculated for unmodified ({\em broken line}) and modified profiles
({\em full lines}) in dependence on the constituent quark mass $M$.\\[35mm]
\end{minipage}
\vspace*{-13mm}

Now let us come back to the inertial parameters and forces we had introduced
to remove center-of-mass motion and to restore good spin and isospin.
Fig.\,9 shows the moment
of inertia and the resulting $\Delta$--nucleon mass-splitting.
The moment of inertia behaves similarly to the r.\,m.\,s.~radius, but
the dominance of the valence quarks is less pronounced. The sea quarks are
responsible for the difference between nucleons and $\Delta$ isobars for
larger masses $M$.
The positive sign of the rotational energy (\ref{EROT}) for $\Delta$ isobars
favors a configuration with a larger moment of inertia.

The difference between nucleon and $\Delta$ mass results mainly from the
rotational energy (\ref{EROT}). In the static case, both particles are
described by a common profile function
$\Theta\sm{(}r\sm{)}$ and have a common moment of inertia
$\cI$.  In this case, the $\Delta$--nucleon mass-splitting is exclusively
given by different rotational energies
\be\label{DNMSst}
\Delta M_{\Delta N}\,=\,\frac{3}{2\cI}.
\ee
Minimizing the corrected soliton energy one gets different meson profiles
resulting in different static energies $E^N$ and
$E^\Delta$, in different CMM corrections $E_{CMM}^N$ and $E_{CMM}^\Delta$,
and in different moments of inertia $\cI_N$ and $\cI_\Delta$.
Now the $\Delta$--nucleon mass-splitting is given by
\be\label{DNMSCORR}
\Delta M_{\Delta N}^{CORR}\,=\,
\left[E^\Delta-E^N\right]-\left[E_{CMM}^\Delta-E_{CMM}^N\right]+
\frac{3}{4}\left(\frac{1}{\cI_\Delta}+\frac{1}{\cI_N}\right).
\ee
The third term can be written in the form of eq.\,(\ref{DNMSst}) introducing
an effective moment of inertia
\be\label{Ieff}
\frac{1}{\cI_{eff}}\,=\,
\frac{1}{2}\left(\frac{1}{\cI_\Delta}+\frac{1}{\cI_N}\right),
\ee
which is the harmonic average of the moments of nucleon and $\Delta$ isobar.
The mass-splitting depends only on the effective moment of inertia
(\ref{Ieff}) but not explicitly on $\cI_N$ and $\cI_\Delta$.
If the first two terms in eq.\,(\ref{DNMSCORR}) are negligible and
the static moment of inertia coincides with the average value (\ref{Ieff})
the mass-splitting for static and corrected meson profiles is the same.
This is obviously the case (fig.\,9, lower part).
The experimental mass-splitting is reproduced for $M\approx$ 430\,MeV.
Here the moments of inertia differ by 10 percent. Nevertheless the mass
splitting is excellently reproduced by the static moment.

Finally we compare the energy corrections themselves. Fig.\,10 illustrates
the differences
between static and self-consistently determined energy corrections.
Rotational corrections are practically the same in both cases.
Merely the CMM energies are noticeably different for $M\gapp$ 400\,MeV.

\hspace*{-38mm}
\begin{minipage}[b]{12.cm}
\vspace*{-20mm}
\mbox{
\psfig{file=fig10a.ps,height=87.15mm,width=14cm,angle=90}}

\vspace*{-52.2mm}

\mbox{
\psfig{file=fig10b.ps,height=100mm,width=14cm,angle=90}}
\end{minipage}
\hfill
\begin{minipage}[b]{5cm}
{\em Figure} 10:
CMM and rotational energy corrections for the nucleon ({\em upper part}  and
for the $\Delta$ isobar ({\em lower part}) calculated with the
corresponding modified profile functions ({\em full lines}).
The {\em broken lines} show the energy corrections for the static profiles.
\\[30mm]
\end{minipage}

\vspace*{-15mm}

\section{Conclusions}

We considered center-of-mass and rotational corrections to solitonic field
configurations of the bosonized Nambu \& Jona-Lasinio model which can be
identified with the nucleon and the $\Delta$ isobar, respectively.
We employed energy corrections which had been derived within the
semiclassical pushing and cranking approaches. The main contribution to the
corrections stems
 from the valence quarks which are confined by the attractive meson field.
We determined modified meson fields by minimizing the static soliton energy
reduced by center-of-mass motion and rotational corrections.
The investigated meson fields were restricted to the chiral circle and to the
hedgehog shape.

We evaluated modified meson and quark fields  as well as expectation values
of several observables in the region 300\,MeV $\lapp M\lapp$ 600\,MeV
of constituent quark masses $M$.
The results illustrate the response of the meson field to the corrections
and quantify their effect on expectation values.
An important effect of the pushing correction is the stabilization of
solitons with light constituent quark masses. CMM corrected solitons
exist for $M\gapp$ 300\,MeV, while
uncorrected solitons are unstable below $M$=350\,MeV.

Despite the big energy corrections meson and quark fields are only
moderately affected. The exclusion of center-of-mass motion narrows the
meson profile and the corresponding quark distribution.
This effect prevents the valence level from diving into the negative-energy
region until very big constituent quark masses.
Rotational corrections affect the asymptotic behavior of the fields at large
radii. They depend on the spin and isospin quantum-numbers and give rise to
differences between
nucleon and $\Delta$ isobar. Moreover they destroy the hedgehog symmetry.
This effect was not considered in the paper.

Both CMM and rotational corrections grow with increasing constituent quark
mass. For $M\gapp$ 500\,MeV the energy corrections reach half of the total
soliton energy and a perturbative treatment seems not to be justified
any more.

In the physically relevant region of small constituent quark masses
(350\,MeV $\lapp M$
\linebreak
$\lapp$ 450\,MeV), the dominating valence-quark picture was confirmed.
The isoscalar r.\,m.\,s.~radius of the nucleon is reduced by a few percent
($\approx$ 3-4 percent for $M$=350\,MeV, $\approx$ 5 percent for
$M$=500\,MeV).
Larger changes were noticed when considering valence, sea or meson
contributions separately.
The slightly different moments of inertia and center-of-mass energies for
nucleons and $\Delta$ isobars do practically not influence the
$\Delta$--nucleon mass-splitting.
The mesonic field energy, which is related to the nuclear $\Sigma$
commutator, is reduced by 20-30 percent.
The general features of the soliton in the bosonized Nambu \& Jona-Lasinio
model are not essentially disturbed by the corrections.\\

The authors wish to acknowledge stimulating discussions with K.\,Goeke,
H.\,Reinhardt, Th.\,Mei{\ss}ner, R.\,Alkhofer, H.\,Weigel, J.\,Berger
and Chr.\,Christov.
The paper was supported by the Bundesministerium f\"ur Forschung
und Technologie (contract 06 DR 107).

\end{document}